\documentclass[referee]{aa} 
\def\note #1]{{\bf #1]}}

\def\al{&\!\!\!\!}
\begin{document}
  \title{Resonant absorption in  dissipative flux tubes}
  \author{H. Safari\inst{1} \and S. Nasiri\inst{1,2} \and K. Karami\inst{1,3,4} \and Y. Sobouti\inst{1}
           }
   \offprints{H. Safari}
    \institute{Institute for Advanced Studies in Basic Sciences, Gava Zang, P.O. Box 45195-1159, Zanjan, Iran
             \and Department of Physics, Zanjan University, Zanjan, Iran
             \and Institute for Theoretical Physics and Mathematics, P.O. Box 19395-5531, Tehran,
             Iran
             \and Department of Physics, University of Kurdistan,
             P.O. Box 66177-15175, Sanandaj, Iran\\
             email: [hsafary, nasiri,karami,sobouti]@iasbs.ac.ir
              }
   \date{Received / Accepted}
\abstract{ We study the resonant absorption of MHD waves in
magnetized flux tubes with a radial density inhomogeneity. Within
the approximation that resistive and viscous processes are
operative in thin layers surrounding the singularities of the MHD
equations, we give the full spectrum of the eigenfrequencies and
damping rates of the MHD quasi modes of the tube. Both surface and
body modes are analyzed. \keywords{Sun -- corona:
magnetohydrodynamics (MHD) -- Sun: magnetic fields -- Sun:
oscillations } }  \maketitle
\section{Introduction} Ionson (1978) was first to suggest  that the resonant absorption
of MHD waves in coronal plasmas  could be a primary mechanism in
coronal heating. Since then, much analytical and numerical work
has been done on the subject. Rae and Roberts (1982) investigated
both eikonal and
 differential equation approaches for the propagation of MHD waves in
 inhomogeneous plasmas. Hollweg (1987a,b) considered a dissipative layer
 of planar geometry to study the resonant absorption of coronal loops.
 Poedts et al. (1989, 1990) developed a finite
element code to elaborate on the resonant absorption of Alfv\'{e}n
waves in circular cylinders.

 Davila (1987) and Steinolfson \&
Davila (1993) did much work on resonant absorption through
resistivity. Ofman et al. (1994) included viscous dissipations in
their analysis and concluded that the  heating rate due to shear
viscosity  is comparable in magnitude to the resistive resonant
heating. Also, they concluded that the heating caused by
compressive viscosity is negligible. Goossens et al. (2002) used
the TRACE data of Ofman \& Aschwanden (2002) to infer the width of
the inhomogeneous layer for 11 coronal loops. Ruderman \& Roberts
(2002) did similar analysis with the data of Nakariakov et al.
(1999). Van Doorsselaere et al. (2004a) used the LEDA code to
study the resistive absorption of  the kink modes of cylindrical
models. They concluded that, when the width of the nonuniform
layer was increased, their numerical results differed  by as much
as 25$\%$ from those obtained with the analytical approximation.
Van Doorsselaere et al. (2004b) investigated the effect of
longitudinal curvature on quasi modes of a typical coronal loop.
They found that the frequencies and damping rates of ideal quasi
modes were
 not  influenced much by the curvature. Andries et al. (2005)
 studied the effect of density stratification on coronal loop
 oscillations, and conclude that longitudinal mode numbers are coupled due to the density
 stratification.

 In the absence of resonance, Edwin \& Roberts (1983) and Roberts
et al. (1984) gave a comprehensive account of the theoretical and
physical concepts related to coronal oscillations.  Karami et al.
(2002; hereafter paper I) studied the full spectrum of MHD modes
of oscillations in zero-$\beta$ magnetic flux tubes with
discontinuous Alfv\'{e}n speeds at the tube's surface. In the
vicinity of singularity, field gradients are large. Recognizing
this, Sakurai et al. (1991a, b) and Goossens et al. (1992, 1995)
developed a method to analyze dissipative processes in such
regimes and to neglect them elsewhere.

Here we combine the two techniques of paper I and of Sakurai et
al. (1991a) to obtain the resonant damping rates for the full
spectrum of the normal modes of magnetic flux tubes.
\section{Equations of motions}
The linearized MHD equations for a zero-$\beta$, but resistive and
viscous, plasma are {\small\begin{equation}
   \frac{\partial\delta\mathbf{v}}{\partial
            t}=\frac{1}{4\pi\rho}\{(\mathbf{\nabla}\times\delta\mathbf{B})\times\mathbf{B}+(\mathbf{\nabla}\times\mathbf{B})\times\delta\mathbf{B}
            \}+\frac{\eta}{\rho}\nabla^2\delta\mathbf{v}\label{1},
   \end{equation}}
   \begin{equation}
 \frac{\partial\delta\mathbf{B}}{\partial
            t}=\mathbf{\nabla}\times(\delta\mathbf{
            v}\times\mathbf{B})+\frac{c^2}{4\pi\sigma}\nabla^2\delta\mathbf{B}\label{2},
   \end{equation}
where $\delta\mathbf{v}$ and $\delta\mathbf{B}$ are the Eulerian
perturbations in the velocity and the magnetic fields; $\rho$,
$\sigma$, $\eta$, and $c$ are the mass density, the electrical
conductivity, the viscosity and the speed of light, respectively.
The simplifying assumptions are:
\begin{itemize}
    \item Under coronal conditions gas pressure is negligible (zero-
    $\beta$).
    \item Density scale heights are much larger than the
    dimensions of flux tube, so that the gravity stratification is
    negligible.
    \item  The tube geometry is a circular cylinder with coordinates
    ($r$,$\phi$,$z$).
    \item There is a constant magnetic field along the $z$ axis,
    $\mathbf{B}=B\hat{z}$.
    \item  The equilibrium is static.
    \item There is no initial steady  flow inside or outside of the tube.
    \item Viscous and resistive coefficients, $\eta$ and $\sigma$
    respectively, are constants.
\end{itemize}
For a variable density, $\rho(r)$, a singularity develops wherever
the local Alfv\'{e}n frequency becomes equal to the global
frequency of the mode. The relevant radial wave number vanishes
and resonant absorption takes place. Let us denote the radius of
the tube by $R$ and a radius beyond which the resonance occurs by
$R_1<R$. The thickness of the inhomogeneous layer, $a=R-R_1$, will
be assumed to be small and will be  arbitrarily taken to be of the
order $R/10$. The choice of density profile is also unimportant.
We will assume two constant densities, $\rho_i$ in $r\leq R_1$ and
$\rho_e<\rho_i$ in $r\geq R$, interconnected with a linearly
varying profile in $R_1\leq r\leq R$.

 In the  remainder of this section the following steps
are taken:
\begin{description}
    \item[a)] In $r<R_1$ and $r>R$, dissipative terms are
neglected. Solutions of Eqs. (\ref{1})
   and (\ref{2}) are obtained as per paper I, and their differences across the inhomogeneous layer
   are calculated.
    \item[b)] In $R_1<r<R$, within which the resonant layer resides, solutions are found by expanding
    Eqs. (\ref{1}) and (\ref{2}) around the singular point, and the jumps across the
    resonant layer are found by the prescriptions of Sakurai et al. (1991a).
    \item[c)] Interior and exterior solutions are connected by requiring
    the jumps in (b) to be equal to the differences in (a). This
    gives an analytical expression for a complex dispersion relation to be solved for the frequencies and the
    damping rates.
\end{description}
\subsection{Interior and exterior solutions}
From paper I, in  the absence of dissipations, all  components of
$\delta\bf{v}$ and the transverse components of $\delta\bf{B}$ are
 expressible in terms of  $\delta B_z$ only. The latter, in turn,
is the solution of a second order differential equation. Thus,
\begin{eqnarray}
\al \al\frac{k^2}{r}\frac{d}{dr}\left[\frac{r}{k^2}\frac{d\delta
B_z}{dr}\right]+(k^2-\frac{m^2}{r^2})\delta B_z=0,\label{dbz}\\
\al\al\delta B_r= \frac{ik_z}{k^2}\frac{d\delta B_z}{dr},~~\delta
B_\phi=-\frac{mk_z}{k^2}\frac{\delta B_z}{r},\label{dbfi}
\end{eqnarray}
\begin{eqnarray}
\delta v_r=-\frac{i}{k^2}\frac{\omega}{B}\frac{d\delta
B_z}{dr},~~~ \delta
v_\phi=\frac{m}{k^2}\frac{\omega}{B}\frac{\delta B_z}{r},~~~
\delta v_z=0,\label{dvr}
\end{eqnarray} where  $k^2=\omega^2/
v_A^2-k_z^2$, and $v_A(r)=B/\sqrt{4\pi\rho(r)}$ is the local
Alfv\'{e}n speed. Here, we have assumed an exponential $\phi,~~z$,
and $t$ dependence, $exp[i(m\phi+k_zz-\omega t)]$ for any
component of $\delta\mathbf{v}$ and $\delta\mathbf{B}$.

In the interior region, $r\leq R_1$, solutions  of Eq. (\ref{dbz})
are{\small
\begin{equation}
\delta B_z=\left\{
\begin{array}{cccc} I_m(|k_i|r),&k_i^2<0,&{\rm ~~surface~waves},&\\
J_m(|k_i|r),&k_i^2>0,&{\rm
body~waves},&\\&k_i^2=\omega^2/v_{A_i}^2-k_z^2,&\end{array}\right.\label{IJ}
 \end{equation}}where $J_m$ and $I_m$ are Bessel
 and modified Bessel functions of the first kind, respectively. In the exterior region, $r\geq R$, the waves should be evanescent.
Solutions are
\begin{equation}
\delta
B_z=K_m(k_er),~~~~~~k_e^2=k_z^2-\omega^2/v_{A_e}^2>0,\label{K}
 \end{equation}
where $K_m$ is the modified Bessel function of  the second kind.
\subsection{Dispersion relation and damping}
From Sakurai et al. (1991a), Goossens et al. (1992, 1995), and
Erd\'{e}lyi et al. (1995), the jump conditions across the boundary
for $\delta B_z$ and $\delta v_r$ are
 \begin{eqnarray}
\al\al[\delta B_z]=0,\label{j1}\\
\al\al[\delta
v_r]~=-\pi\tilde{\omega}\frac{1}{|\Delta|}\frac{m^2}{\rho(r_A)r_A^2}B_z
\delta B_z,\label{j2}
\end{eqnarray}
 where $R_1<r_A<R$~ is the radius at which the singularity occurs
and $k^2(r_A)=0$, $\tilde{\omega}=\omega+i\gamma$, and
 $\Delta=-B^2\frac{d}{dr}(\frac{k^2}{\rho})|_{r_A}$. Substituting the
 fields of Eqs. (\ref{dvr}), (\ref{IJ}) and (\ref{K}) in jump
 conditions and eliminating the arbitrary amplitudes of the waves,
 as foreseen initially inside and outside of the boundary layer, gives
 the dispersion relation
\begin{eqnarray}
d_0(\tilde{\omega},m)+d_1(\tilde{\omega},m)=0,\label{disp1}
 \end{eqnarray}
 where
\begin{eqnarray}
&&d_0(\tilde{\omega},m)=-\frac{1}{k_e}\frac{K_m'(|k_e|R)}{K_m(|k_e|R)}+
\frac{1}{k_i}\frac{J_m'(|k_i|R_1)}{J_m(|k_i|R_1)}\label{d0},\\
&&d_1(\tilde{\omega},m)=-i\pi
\frac{1}{|\Delta|}\frac{m^2}{\rho(r_A)r_A^2}\label{d1}.
\end{eqnarray}
In principle $\tilde{\omega}=\omega+i\gamma$ is expected to be found
as a solution of Eqs. (\ref{disp1}-\ref{d1}). In particular for
$\gamma\ll \omega$, Eq. (\ref{disp1}) can be expanded about $\omega$
to give
\begin{eqnarray} \gamma=-\left.\frac{\rm{Im}(d_1(\tilde{\omega},m))}{\partial d_0(\tilde{\omega},m)/\partial\tilde{\omega}}
\right|_{\tilde{\omega}=\omega}\label{damp}.\end{eqnarray} The
results for surface waves are the same  as   Eqs. (\ref{disp1})-
(\ref{damp}), except that $J_m$ is replaced by $I_m$ everywhere.
This completes the formal solutions of $\tilde{\omega}$ and
$\gamma$. Further analytical progress is still possible when the
inhomogeneous layer is thin enough.
\section{Thin boundary approximation}
Here we assume  $(R-R_1)/R=a\ll1$. Equation (\ref{disp1}) reduces
to $d_0\approx0$. For body waves, the latter is studied in ample
detail in paper I. There, a trio of wave numbers $(n,m,l)$,
corresponding to $r$, $\phi$ and $z$ directions, is assigned to
each mode. From Eq. (\ref{damp}) the corresponding damping rate
becomes
\begin{eqnarray}
\gamma_{nml}&=&-\left\{\frac{\pi m^2}{\omega_{nml}^2
(\rho_i-\rho_e)}\frac{a}{R^2}\right\}/\frac{d}{d\omega}\left\{
\frac{1}{k_e}\frac{K_m'(|k_e|R)}{K_m(|k_e|R)}\right.\nonumber\\&&\left.
-\frac{1}{k_i}\frac{J_m'(|k_i|R)}{J_m(|k_i|R)}\right\}.
\label{gam}
\end{eqnarray}
Again the results for surface waves are the same as those for body
waves except that $J_m$ is replaced by $I_m$. The two waves
exhibit differences, for $J_m$ and $I_m$ behave differently at the
boundary. We also note that each surface mode is designated by
only two wave numbers, $(m,l)$ corresponding to $\phi$ and $z$
directions, respectively.

 For $m=0$, resonant absorption  does not take place, because the jumps
in $\delta B_z$ and $\delta v_r$ vanish and Eqs. (\ref{disp1}-
\ref{damp}) are not valid anymore.

\section{Numerical results}
As typical parameters for a coronal loop, we adopt radius = $10^{3}$
km, length = $10^{5}$ km, $\rho_{i}=2\times 10^{-14}$ gr cm$^{-3}$,
$\rho_{e}/\rho_{i}=0.1$, $B=100$ G. For these parameters one finds
$v_{A_{i}}=2000$ km s$^{-1}$, $v_{A_{e}}=6400$ km s$^{-1}$ and
$\omega_{A}=2$ rad s$^{-1}$.

\begin{center}
\begin{figure}[h]
\includegraphics{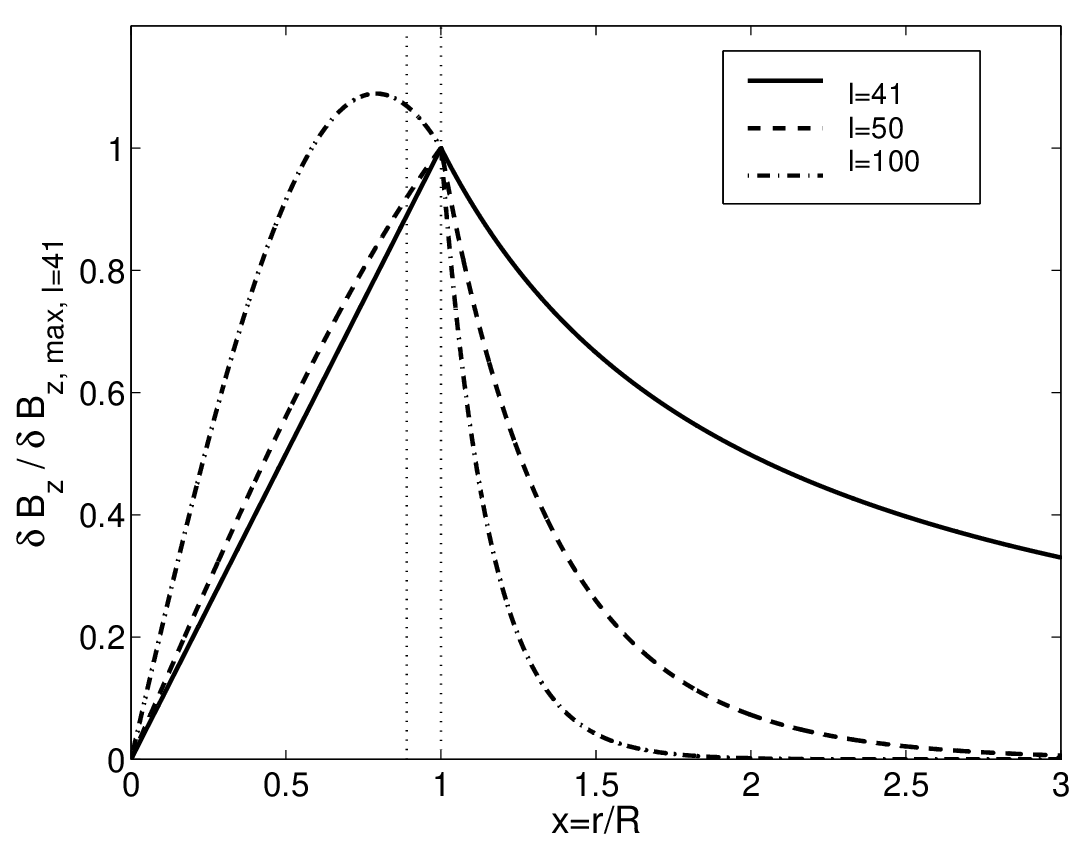}
      \vspace{6.cm}
      \caption[]{Magnetic component $\delta B_z/{\delta B_z}_{max,~l=41}$,
      for body modes, where $n=1$, $m=1$ (kink), and $l=41, 50, 100$. Slopes are
      discontinuous at $x=1$.
       The maximum wave amplitude shifts towards the tube axis as $l$ increases.}
         \label{boydm14050100}
   \end{figure}
   \end{center}

 In Fig. \ref{boydm14050100} the magnetic field component,
$\delta B_z(r)$,  for  $n=1$, $m=1$ and $l=41,~50,~100$, are
plotted versus  $x=r/R$. They are normalized to  $\max (\delta
B_z, l=41)$.  The segment of the plot in $0<x<1$ is from the
interior solutions of Eq. (\ref{IJ}), while the segment in $x>1$
is from the exterior solutions of Eq. (\ref{K}).  Slopes are
discontinuous at $x=1$. Actually, the correct location of the
discontinuities is the point of singularity, $r_A$. Its appearance
 at $x=1$ is an artifact caused by extrapolating the interior
 solutions to cover the boundary layer, rather than
 using the exact solutions there. As $l$ increases, the
maximum wave  amplitude moves towards the tube axis and away from
the inhomogeneous layer.

\begin{center}
\begin{figure}[h]
\includegraphics{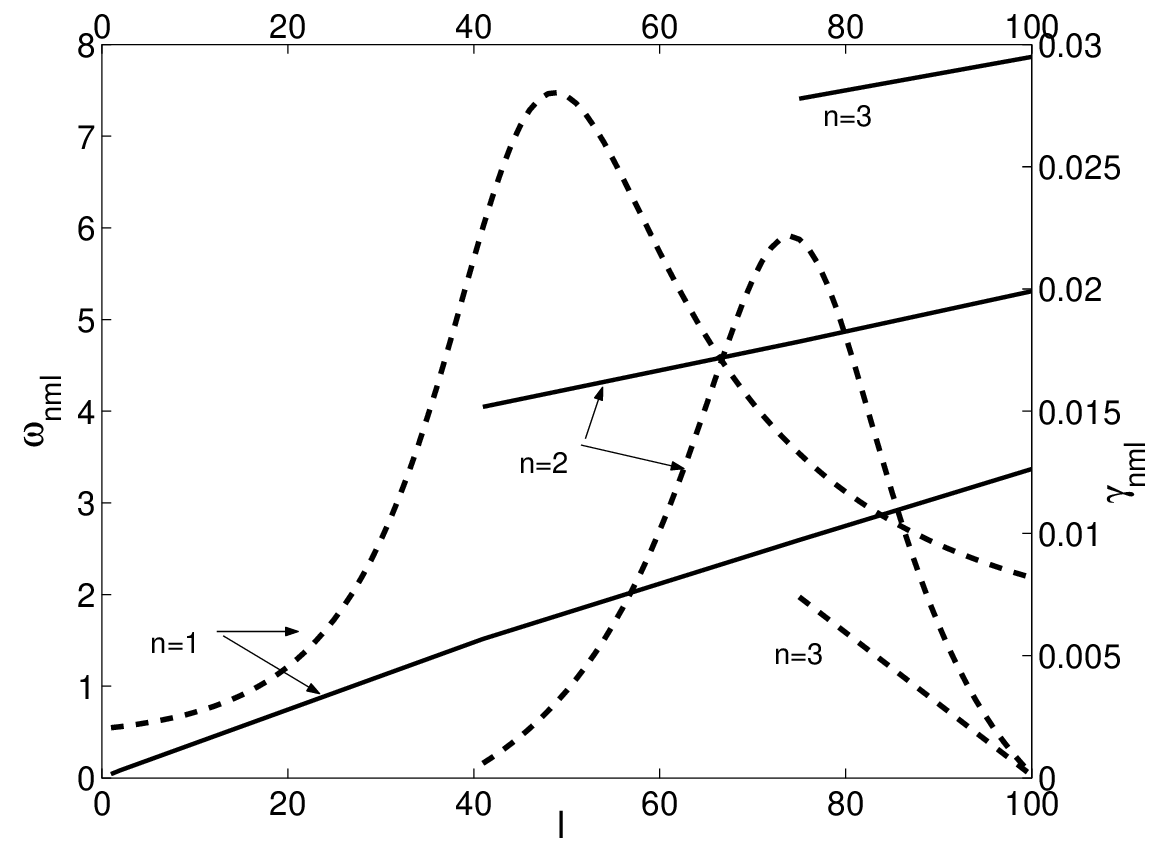}
      \vspace{6.4cm}
      \caption[]{Body modes: $\omega_{nml}$ (solid lines) and $\gamma_{nml}$
        (dashed lines)  versus $l$
      for $m=1$ (kink) modes. Auxiliary parameters are:  the tube radius
$10^{3}$ km, the tube length  $10^{5}$ km, $B=100$ G,
$\rho_{i}=2\times 10^{-14}$ grcm$^{-3}$, $\rho_{e}/\rho_{i}=0.1$.
Frequencies are in units of the interior Alfv\'{e}n frequency,
$\omega_{A}=2$ rad s$^{-1}$.}
         \label{bm1}
   \end{figure}
\end{center}

In Fig. \ref{bm1}, the frequencies and the damping rates are
plotted versus $l$. The frequencies increase with increasing  $n$
and/or $l$. For $n=1$ and $2$, damping rates exhibit maxima at
$l\approx50$ and $70$, respectively. For $n=3$ there is only a
declining branch towards higher $l$ values. Our explanation for
the occurrence of maxima in damping rates is the localization of
the maximum amplitude of a wave within the resonant layer, where
the dissipation is expected to be the highest. As Fig.
\ref{boydm14050100} shows, this maximum moves away from that layer
at both lower and higher $l$ values.

\begin{center}
\begin{figure}[h]
\includegraphics{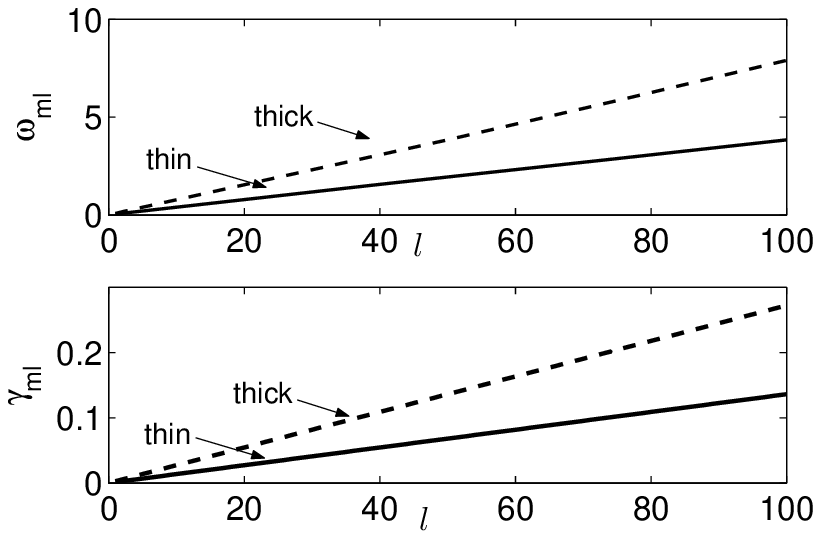}
      \vspace{5.cm}
          \caption[]{Surface modes:  $\omega_{ml}$ and $\gamma_{ml}$versus $l$ for $m=1$,
           $R/L=0.01$ (Solid line) \& $0.02$ (Dashed line). Auxiliary parameters as in
           Fig. \ref{bm1}.} \label{sm1}
   \end{figure}
   \end{center}

\begin{center}
\begin{figure}[h]
\includegraphics{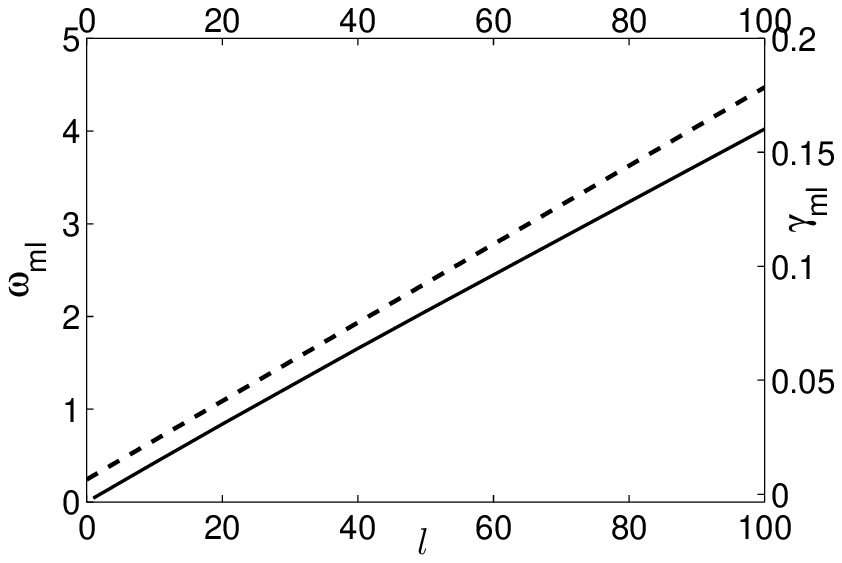}
      \vspace{3.3cm}
          \caption[]{Surface modes: $\omega_{ml}$ and
      $\gamma_{ml}$ for $m=2$ and $R/L=0.01$.
      Auxiliary parameters as in  Fig. \ref{bm1}.} \label{sm2}
   \end{figure}
   \end{center}

For surface waves, specified by two mode numbers $(m,l)$,
$\omega_{ml}$ and $\gamma_{ml}$ are plotted in Figs. \ref{sm1} and
\ref{sm2}. Both frequencies and damping rates show a monotonous
increase with $l$. Two tube radii, $R/L=0.01,~\&~0.02$, are
considered here. The frequencies and damping rates of the thicker
tube are almost double those of the thinner tube. This is
expected, because in the limit of thin tubes, both $\omega$ and
$\gamma$ are proportional to $R/L$  (see, e. g., Van Doorsselaere
et al. 2004). The frequencies of the surface waves of Fig.
\ref{sm1} and the $n=1$ body frequencies of Fig. \ref{bm1}  both
behave similarly. This is because of the similar behavior of
$I_{m}(kr)$ and $J_m(kr)$ at small arguments. In physical terms
this means, at least in the thin tube approximation, the surface
and the $r$- fundamental body waves contribute in equal manner to
the heating mechanism; they both dump the bulk of their energies
near the tube surface, and perhaps both are excited with
comparable amplitudes and energies. The $\omega/2\pi\gamma$ is the
number of oscillations  taking place before a wave is completely
attenuated. For thick and thin tubes,
 and $1\leq l\leq 10$, this number  is about $1.7$ and $8$,
respectively. An observed value of  Nakariakov et al. (1999) from
TRACE data is about $3.39$. Those of Goossens et al. (2002) from
the same source range from $1.01$- $3.21$ corresponding, according
to the authors, to various  values of $R/L$, $a/R$, and
$\rho_e/\rho_i$.

\section{Summary}  We studied the MHD quasi modes of coronal
loops.  On the assumption that ohmic and viscous dissipations are
operative within a thin boundary layer, we obtained  an analytic
dispersion relation and solved it numerically for the mode
frequencies and the damping rates. For realistic values of the
initial parameters  thickness- to- height ratio of the loop, the
density contrast with the background medium, and the equilibrium
magnetic field- our numerical values agree with those obtained
from observations. As the longitudinal wave number increases, the
maximum amplitude of the body eigenmodes shifts away from the
resonant layer and causes a decrease in damping rates. Its
behavior with increasing radial wave number is not, however, all
that straightforward.  In additional, we have shown that body and
surface modes may contribute equally to the heating of coronal
loops.
\begin{acknowledgements}
The authors wish to thank Professors  Robert Erd\'{e}lyi and,
Marcel Goossens for valuable consultations. Elaborate and
meticulous comments of the  referee has significantly improved the
content and the presentation of the paper. This work was supported
by the Institute for Advanced Studies in Basic Sciences (IASBS),
Zanjan.
\end{acknowledgements}

\clearpage
\end{document}